\documentclass[floats, aps, showpacs, twocolumn, superscriptaddress, prl,
    reprint, floatfix]{revtex4-1}

\usepackage{graphicx}
\usepackage{amsmath, amssymb, amstext, amsthm, amsfonts}
\usepackage{bbm}
\usepackage[multidot]{grffile}
\usepackage{hyperref}

\newcommand{\ue}{\ensuremath{\text{e}}}

\newcommand{\ud}{\ensuremath{\text{d}}}
\newcommand{\saddle}{\ensuremath{\Gamma_{\text{s}}}}
\newcommand{\hsaddle}{\ensuremath{\Gamma^h_{\text{s}}}}

\newcommand{\Xbwd}{\ensuremath{\Gamma_{\text{b}}}}
\newcommand{\exit}[1][0]{\ensuremath{\mathcal{E}^h_{#1}}}
\newcommand{\exitn}{\ensuremath{\mathcal{E}^h_{n}}}
\newcommand{\munat}{\ensuremath{\mu_{\text{nat}}}}
\newcommand{\mugh}{\ensuremath{\mu_{\gamma}^h}}
\newcommand{\mugamma}{\ensuremath{\mu_{\gamma}}}

\newcommand{\gnat}{\ensuremath{\gamma_{\text{nat}}}}

\newcommand{\Map}{\ensuremath{M}}

\newcommand{\QMap}{\ensuremath{U}}
\newcommand{\QMapcls}{\ensuremath{U}_\text{cl}}
\newcommand{\tcsmax}{\ensuremath{m_h}}
\newcommand{\sdist}[1][x]{\ensuremath{t_h(#1)}}

\newcommand{\Qavg}[1][\ensuremath{\gamma}]{
	\ensuremath{\langle \mathcal{H} \rangle_{#1}}}

\newcommand{\hone}{\ensuremath{(i)}}		
\newcommand{\htwo}{\ensuremath{(ii)}}

\newcommand{\HIDDEN}[1]{}

\usepackage{color}
\begin{document}

\title{Resonance eigenfunction hypothesis for chaotic systems}

\newcommand{\affilTUD}{
    Technische Universit\"at Dresden,
    Institut f\"ur Theoretische Physik and Center for Dynamics,
    01062 Dresden, Germany}
\newcommand{\affilMPI}{
    Max-Planck-Institut f\"ur Physik komplexer Systeme,
    N\"othnitzer Stra\ss{}e 38, 01187 Dresden, Germany}

\author{Konstantin Clau\ss}
\affiliation{\affilTUD}

\author{Martin J.\ K\"orber}
\affiliation{\affilTUD}

\author{Arnd B\"acker}
\affiliation{\affilTUD}
\affiliation{\affilMPI}

\author{Roland Ketzmerick}
\affiliation{\affilTUD}
\affiliation{\affilMPI}

\date{\today}

\begin{abstract}
   A hypothesis about the average phase-space distribution of 
   resonance eigenfunctions in chaotic systems
   with escape through an opening is proposed.
   Eigenfunctions with decay rate $\gamma$ are described by a classical 
   measure that
   $\hone$ is conditionally invariant with classical decay rate $\gamma$ and
   $\htwo$ is uniformly distributed on sets with the same temporal
   distance to the  quantum resolved chaotic saddle.
   This explains the localization of fast-decaying resonance eigenfunctions
   classically. 
   It is found to occur in the phase-space region having the 
   largest distance to the chaotic saddle.
   We discuss the dependence on the decay rate $\gamma$ and the
   semiclassical limit.
   The hypothesis is numerically demonstrated for the standard map. 
\end{abstract}

\maketitle

\noindent

%
\emph{Introduction.}---%
Eigenvalue spectra and the structure of eigenfunctions are the key to
understanding any quantum system. %
Universal properties are usually expected
for quantum systems with chaotic classical dynamics.
For closed systems the \emph{statistics of eigenvalues} follows random matrix 
theory \cite{BohGiaSch1984,Ber1985,SieRic2001,MueHeuBraHaaAlt2004}, and
the \emph{structure of eigenfunctions} is described by the 
{semiclassical eigenfunction hypothesis} \cite{Vor1979,Ber1977b,Ber1983}.
It states that eigenfunctions are concentrated on those regions explored by
typical classical orbits.
If the dynamics is ergodic, this is proven by the quantum ergodicity theorem
\cite{Shn1974,CdV1985,Zel1987,ZelZwo1996,Bie2001,NonVor1998},
showing that almost all eigenfunctions 
converge to the uniform distribution on the energy shell
in phase space \cite{BaeSchSti1998}.
These fundamental results for single particle quantum chaos recently had strong
impact in many-body systems, e.g.\ for thermalization 
\cite{Sre1994,AleKafPolRig2016}.

Experimentally one often deals with chaotic scattering
systems \cite{Gas2014b}, which appear
in many fields of physics, such as
nuclear reactions \cite{MitRicWei2010},
microwave resonators \cite{Sto1999},
acoustics \cite{TanSoe2007},
quantum dots \cite{Jal2016}, and optical microcavities \cite{CaoWie2015}. %
Thus the counterparts of the fundamental results of
closed systems are desired for scattering systems.
This has been achieved for the \emph{statistics of resonances}
\cite{HaaIzrLehSahSom1992,FyoSom1997,Alh2000,%
      JacSchBee2003,FyoSav2012,KumDieGuhRic2017,%
      Sjo1990,Lin2002,LuSriZwo2003,SchTwo2004,
      NonZwo2005,She2008,RamPraBorFar2009,KopSch2010,KoeMicBaeKet2013}.
In particular the fractal Weyl law 
\cite{Sjo1990,Lin2002,LuSriZwo2003,SchTwo2004,NonZwo2005,She2008,%
      RamPraBorFar2009,KopSch2010,KoeMicBaeKet2013}
relates the growth rate of the number of long-lived resonances to the
fractal dimension of the chaotic saddle of the classical dynamics.
For the \emph{structure of resonance eigenfunctions}
some aspects have been studied,
e.g.\ for open billiards %
\cite{IshSaiSadBer2001,
    KimBarStoBir2002,
    BaeManHucKet2002,
    WeiRotBur2005,
    WeiBarKuhPolSch2014},
optical microcavities %
\cite{GmaCapNarNoeStoFaiSivCho1998,
    LeeRimRyuKwoChoKim2004,
    WieHen2008,%
    ShiHarFukHenSasNar2010,%
    ShiWieCao2011,%
    HarShi2015},
potential systems \cite{RamPraBorFar2009},
and maps
\cite{CasMasShe1999b,
    SchTwo2004,
    KeaNovPraSie2006,
    NonRub2007,
    ErmCarSar2009,
    LipRyuLeeKim2012,
    CarWisErmBenBor2013,
    SchAlt2015,
    KoeBaeKet2015}.
However, there exists no analogue to the semiclassical eigenfunction 
hypothesis for scattering systems. %
This fundamental open problem of the structure of resonance eigenfunctions
is addressed in this paper.

\begin{figure}[b!]
    \includegraphics{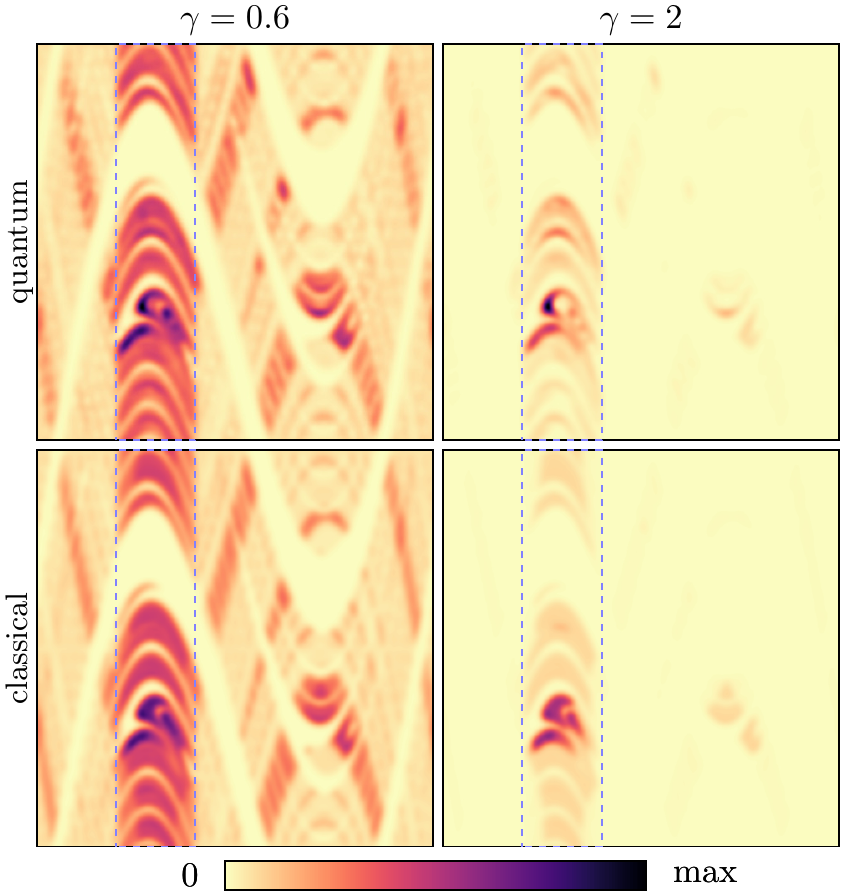}
    \caption{Average Husimi phase-space
        distribution of resonance eigenfunctions (top)
        compared to constructed classical measures $\mugh$ (bottom)
        with decay rates $\gamma = 0.6$ (left) and $\gamma = 2$ (right)
        for $h = 1/1000$.
        Chaotic standard map with $\kappa = 10$ on phase space
        $\Gamma = [0,1)\times[0,1)$ with
        opening $\Omega = [0.2, 0.4]\times[0, 1)$ (blue dashed line).
        Colormap with fixed maximum for each $\gamma$.
        Prominent localization for $\gamma = 2$ and overall
        quantum-classical agreement.
    }
    \label{FIG:title}
\end{figure}

\begin{figure}[b!]
    \includegraphics[scale=1.]{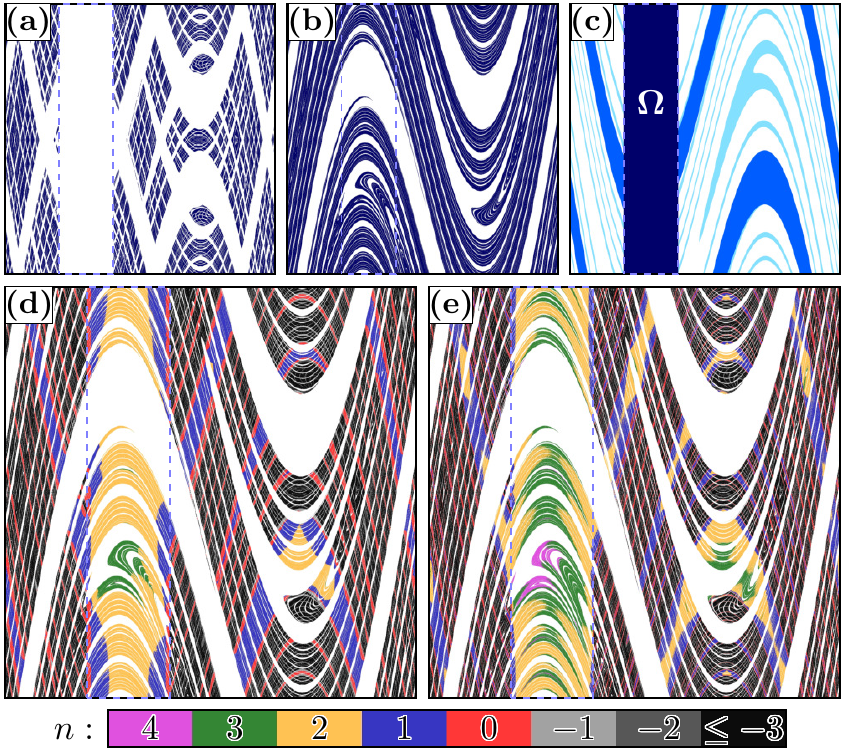}
    \caption{Classical sets for the considered standard map.
        (a) Chaotic saddle $\saddle$, (b) backward trapped set $\Xbwd$,
        (c) opening $\Omega$ and preimages
        $\Map^{-1}(\Omega), \Map^{-2}(\Omega)$ (from dark to light),
        (d, e) 
        partition of the backward trapped set $\Xbwd$ with colored
        sets $\exit[n]$ with integer saddle distance $n \leq \tcsmax$ for
        (d) $h = 1/1000$ with $\tcsmax = 3$ and
        (e) $h = 1/16000$ with $\tcsmax = 4$.
        Regions with $n \leq 0$ are within the $h$-resolved saddle 
        $\hsaddle$.%
    }
    \label{FIG:TIMES}
\end{figure}

For simplicity of the presentation we focus in the rest of the paper on
time-discrete maps %
with chaotic dynamics and escape through an opening. 
The resulting discussion is straightforwardly generalized to autonomous systems
like the paradigmatic three-disk scattering system or the
H\'{e}non-Heiles potential.
Maps arise naturally, e.g.,
from a Poincar\'{e} section in autonomous systems or
from a stroboscopic Poincar\'{e} section in time-periodically 
driven systems.
Quantizing such a map yields a subunitary propagator %
whose non-orthogonal, right eigenfunctions %
have varying decay rates $\gamma$.
Note that these eigenfunctions extend into the opening
for the chosen ordering of escape before the mapping, see Fig.~\ref{FIG:title}.
Surprisingly, there occurs a localization of their average
phase-space distribution within the opening.
This localization is more prominent for resonance eigenfunctions
with large decay rates $\gamma$,
as visualized for the standard map (introduced below)
in Fig.~\ref{FIG:title} (top).
Thus, the following questions arise:
What is the origin of this localization?
What distinguishes the phase-space region of localization?
More generally, is this effect caused by quantum interference
(like dynamical localization \cite{Fis2010} or
scarring due to periodic orbits \cite{Hel1984})
or by properties of the classical dynamics?
Before answering these questions, let us briefly introduce the classical and
quantum mechanical background.
Classically, for chaotic dynamics of a map with escape
almost all points on phase-space $\Gamma$
will be mapped into the opening $\Omega$ eventually and thus escape 
\cite{LaiTel2011}.
Only a set of measure zero does not leave the system under
forward and backward iteration. This invariant set usually is a
fractal and is called the \emph{chaotic saddle} $\saddle$,
see Fig.~\ref{FIG:TIMES}(a).
Its unstable manifold consists of
points approaching $\saddle$ under the inverse map
and is therefore called the \emph{backward-trapped set} $\Xbwd$,
see Fig.~\ref{FIG:TIMES}(b).
Generic initial phase-space distributions
asymptotically converge to the uniform distribution on $\Xbwd$, the so-called
\emph{natural measure} $\munat$, with corresponding decay rate $\gnat$
\cite{PiaYor1979,KanGra1985,Tel1987,LopMar1996,DemYou2006,AltPorTel2013}.
Quantum mechanically, the support of resonance eigenfunctions is given by
the backward trapped set $\Xbwd$ \cite{CasMasShe1999b,KeaNovPraSie2006}.
Furthermore, long-lived eigenfunctions with decay rates $\gamma\approx\gnat$
are distributed as the natural measure $\munat$ on phase space
\cite{CasMasShe1999b}, which corresponds to the
steady-state distribution in the context of optical microcavities
\cite{LeeRimRyuKwoChoKim2004}.
There are a few supersharp resonances with
$\gamma$ significantly smaller than $\gnat$ \cite{Nov2012}.
Instead, we focus on the large number of
shorter-lived eigenfunctions ($\gamma > \gnat$).
For their integrated weight on $\Omega$ and on each of its preimages
the dependence on the decay rate $\gamma$ was derived in
reference \cite{KeaNovPraSie2006}.
This concept is generalized by so-called conditionally invariant measures
\cite{PiaYor1979,LopMar1996,DemYou2006,NonRub2007}.
Recently, we suggested a specific conditionally invariant measure
proportional to $\munat$
on the opening $\Omega$, describing classically the weight of eigenfunctions
on either side of a partial barrier \cite{KoeBaeKet2015}.
None of these results, however, explains the observed localization
phenomenon. %

In this paper we propose a hypothesis for resonance eigenfunctions
in chaotic
systems predicting their average phase-space distribution.
The hypothesis defines a conditionally invariant measure
of the classical system
for given decay rate $\gamma$ and effective Planck's constant $h$.
It gives a classical explanation  for the %
localization of resonance eigenfunctions in those phase-space regions
having the largest distance to the chaotic saddle.
This is demonstrated in Fig.~\ref{FIG:title} for the chaotic standard map.
We discuss the dependence on $\gamma$ and $h$,  
and briefly speculate about the semiclassical limit.

\emph{Resonance eigenfunction hypothesis.}---%
We postulate that in chaotic systems with escape through an opening
the average phase-space distribution of resonance eigenfunctions
with decay rate $\gamma$ for effective Planck's constant $h$
is described by a measure that
$\hone$ is conditionally invariant with decay rate $\gamma$ and
$\htwo$ is uniformly distributed on sets with the same temporal distance
to the $h$-resolved chaotic saddle.

Combining both properties yields a measure
\begin{align}
    \mugh(A) = \frac{1}{\mathcal{N}} \int_A
    \ue^{\sdist\cdot(\gamma - \gnat)}\,\ud\munat(x),
    \label{EQ:mugh0}
\end{align}
for all $A\subset \Gamma$ with normalization constant $\mathcal{N}$.
Here the temporal \emph{saddle distance} $\sdist\in\mathbb{R}$ fulfills
\begin{align}
    \sdist[\Map^{-1}(x)] = \sdist - 1   \label{EQ:dist}
\end{align}
for almost all $x\in\Xbwd$, i.e.\ each backward iteration of the map 
$\Map$ on $\Xbwd$ reduces the saddle distance by one.
An important implication of Eq.~\eqref{EQ:mugh0} is that %
$\mugh$ is enhanced with increasing $\gamma > \gnat$ in those
regions of $\Xbwd$ having the largest saddle distance,
due to the exponential factor.
These regions must be in the opening $\Omega$, which is easily shown by
contradiction. %
Thus the hypothesis leads to a classical prediction
for the localization of resonance eigenfunctions in chaotic systems.

We will now discuss properties $\hone$ and $\htwo$ in more detail.
A measure $\mu$ is called conditionally invariant with decay rate $\gamma$
under a map $\Map$ with escape through an opening,
if it is invariant under time evolution up to an overall decay,
\begin{align}
    \mu(\Map^{-1}(A)) = \ue^{-\gamma}\,\mu(A), \label{EQ:ci}
\end{align}
for all  $A\subset \Gamma$ \cite{PiaYor1979,LopMar1996,DemYou2006}.
Equation \eqref{EQ:ci}  
states that the set $M^{-1}(A)$, which consists of points
that are mapped onto $A$,
has a measure that is smaller by a factor $\ue^{-\gamma}$
than the measure of $A$.
The support of conditionally invariant measures is the backward trapped set 
$\Xbwd$.
The most important of these measures is the
natural measure $\munat$ with decay rate $\gnat$ \cite{LopMar1996,DemYou2006}.
This measure is uniformly distributed on the backward trapped set $\Xbwd$.
We stress that for any decay rate $\gamma$ there are infinitely many different
conditionally invariant measures \cite{DemYou2006,NonRub2007}.
So far it is unknown, if any of these classical measures corresponds to
resonance eigenfunctions %
with arbitrary decay rates.

Property $\htwo$ selects a specific class of measures
which are uniformly distributed on subsets of $\Xbwd$. %
Uniform distribution with respect to $\Xbwd$ (the support 
of conditionally invariant measures) is equivalent to
proportionality to the natural measure,
explaining the appearance of $\munat$ in Eq.~\eqref{EQ:mugh0}.
In analogy to quantum ergodicity for closed systems
it is reasonable to consider for resonance eigenfunctions a
uniform distribution on $\saddle$,
as classically this is an invariant set with chaotic dynamics.
The quantum mechanical uncertainty relation, however,
implies a finite phase-space 
resolution $h$ replacing $\saddle$ by a quantum resolved saddle $\hsaddle$.
It is desirable to combine the assumption of uniformity on the saddle,
the finite quantum resolution, and conditional invariance.
This is achieved by introducing a temporal distance $\sdist$ to
the quantum resolved saddle $\hsaddle$
for all $x\in\Xbwd$ and assuming uniformity on all sets with
the same temporal distance.
The resulting measures $\mugh$, Eq.~\eqref{EQ:mugh0},
are conditionally invariant according to Eq.~\eqref{EQ:ci}
as can be shown using Eq.~\eqref{EQ:dist}.

For the saddle distance $\sdist$ we now provide a conceptually
and numerically simple implementation.
For this we consider
as a convenient definition of $\hsaddle$ %
a symmetric surrounding of $\saddle$,\
$\hsaddle = \{x\in\Gamma : d(x, \saddle) \leq \sqrt{\hbar/2} \}$,
with Euclidean distance $d$ smaller than the width of coherent states.
We define an \emph{integer saddle distance} $n\in\mathbb{Z}$
for $x\in\Xbwd$ as
the number of backward steps to enter the $h$-resolved saddle, 
\begin{align}
\sdist = n \ \, \Leftrightarrow \ \,
\Map^{-n}(x) \in \hsaddle,
\end{align}
with $\Map^{-i}(x) \notin \hsaddle$ for all $i < n$.
For points inside of $\hsaddle$ this leads to $n\leq 0$.
Defining $\exitn := \{x\in\Xbwd : \sdist = n\}$ as the sets with
integer saddle distance $n$ we obtain a partition of $\Xbwd$ with
$\exitn = \Map^n\exit$.
There is a maximal saddle distance $\tcsmax$,
and consequently the regions $\exit[n]$ with $n > \tcsmax$ are empty sets.
With this Eq.~\eqref{EQ:mugh0} simplifies to
\begin{align}
\mugh(A) &= \frac{1}{\mathcal{N}}\sum_{n = -\infty}^{\tcsmax} 
\ue^{n(\gamma - 
\gnat)}
\munat(A\cap \exitn), \label{EQ:mugh}
\end{align}
for all $A\subset \Gamma$, which will be applied in the following.
%

\emph{Example system.}---%
Throughout this paper we use the paradigmatic example of the
standard map \cite{Chi1979} %
in its symmetric form
$(q, p) \mapsto (q + p^\ast, p^\ast + v(q + p^\ast))$ with $p^\ast = p + v(q)$
and $v(q) = (\kappa/4\pi) \sin(2\pi q)$,
considered on the torus %
$q\in[0,1)$, $p\in[0,1)$ with periodic boundary conditions.
We consider a kicking strength
$\kappa = 10$ to ensure a fully chaotic phase space.
The opening is chosen as a vertical strip $\Omega$, such that
$ q\in [0.2, 0.4]$ and $p\in [0, 1)$, see Fig.~\ref{FIG:title}.
Position and size of $\Omega$ determine the classical decay rate
$\gnat \approx 0.21$ of the natural measure $\munat$.

We consider the Floquet quantization $\QMapcls$
\cite{BerBalTabVor1979,ChaShi1986}
of the closed map on a Hilbert space of dimension $1/h$ %
with effective Planck's constant $h$.
The quantum map is opened as 
$\QMap = \QMapcls\cdot (\mathbbm{1} - P_{\Omega})$
with projector $P_\Omega$ on the opening \cite{BorGuaShe1991}. %
The eigenvalue problem of this subunitary
propagator, $U\psi = \lambda\,\psi$, leads to eigenvalues with
modulus less than unity, $|\lambda|^2 \equiv \ue^{-\gamma} < 1$.
The decay rate $\gamma$ characterizes the time evolution of the
corresponding resonance eigenfunction $\psi$.
There is a broad distribution of decay rates $\gamma$ 
\cite{BorGuaShe1991,SchTwo2004}.
We compute the Husimi phase-space distribution
$\mathcal{H}(q, p) = 1/h\ |\langle q, p | \psi\rangle|^2$
for each eigenfunction $\psi$
by taking the overlap with symmetric coherent states $|q, p\rangle$
centered at $(q, p)\in\Gamma$.

While Husimi distributions $\mathcal{H}$ of individual
resonance eigenfunctions show strong
quantum fluctuations, %
we want to explain their average behavior.
Therefore we calculate the average Husimi distribution
$\Qavg[{\gamma}]$,
where the average is taken over eigenfunctions from the interval
$[{\gamma}\cdot c, {\gamma}/c]$ around some $\gamma$-value of interest
with constant $c = 0.95$.
We improve this averaging by increasing the number of contributing
Husimi distributions in two ways:
First, we vary the Bloch phase $\theta_p \in \{0.04, 0.08, \dots, 0.96\}$
of the quantization $U$.
Secondly, the inverse Planck's constant is varied in
$\{0.94, 0.96, 0.98, 1, 1.02, 1.04, 1.06\}\cdot h^{-1}$ for $h=1/1000$.

Classical measures $\mugh$ are obtained as follows. 
Using the sprinkler method \cite{LaiTel2011}
we approximate the chaotic saddle $\saddle$ as a point set with
more than $10^7$ points not leaving the system under ten forward and
backward time steps, see Fig.~\ref{FIG:TIMES}(a).
Tenfold forward iteration of this set
gives an approximation of $\Xbwd$, see Fig.~\ref{FIG:TIMES}(b).
The uniform distribution on this point set approximates
$\munat$ which is used in Eq.~\eqref{EQ:mugh}.
We partition $\Xbwd$ into sets $\exitn$
by determining the integer saddle distance $n$ for each $x\in \Xbwd$,
such that
$d(\Map^{-n}(x), \saddle) \leq \sqrt{\hbar/2}$ and
$d(\Map^{-n + 1}(x), \saddle) > \sqrt{\hbar/2}$,
shown in Figs.~\ref{FIG:TIMES}(d) and (e) for two values of $h$. 
Note that the region with maximal saddle distance $\tcsmax$
is similar for both considered $h$.
The saddle distance $n$ varies for points on $\Xbwd$ and in particular
on the opening $\Omega$ for two reasons:
the geometric distance along the manifold to the quantum
resolved saddle $\hsaddle$ and the variation of the local stretching,
i.e.\ finite time Lyapunov exponents.
In order to construct $\mugh$,
we assign to each $x\in\Xbwd$ a weight  $\ue^{n(\gamma - \gnat)}$
according to the factor in Eq.~\eqref{EQ:mugh}.
Integrating these weights over grid cells with chosen
resolution $800 \times 800$ and normalizing
we obtain a phase-space density numerically approximating $\mugh$.
%

\emph{Comparison.}---%
In Fig.~\ref{FIG:title} we show the average
phase-space distributions $\Qavg[\gamma]$ for $\gamma = 0.6$ and $\gamma = 2$
for $h=1/1000$.
Because $\mathcal{H}(q, p)$ is the expectation value of the projector on a 
coherent state $|q, p\rangle$, we compute
the classical analogue.
This is obtained by a convolution of the constructed measures $\mugh$
with a Gaussian of the same width
as the coherent state, i.e.\ with standard deviation $\sqrt{\hbar/2}$.
This allows for quantum-classical comparison on the phase space.
Overall we observe very good agreement
concerning the support of the distributions,
their weight on the opening $\Omega$,
and their localization within $\Omega$.

The Husimi distributions show the following features:
First, they %
are supported by the smoothed backward trapped set.
Secondly, one observes that their density on the opening $\Omega$ is larger
than on its surrounding.
The other stripes with larger density (than their surrounding)
fall on the preimages $\Map^{-1}(\Omega)$ and $\Map^{-2}(\Omega)$,
shown in Fig.~\ref{FIG:TIMES}(c).
Thirdly and most importantly, the Husimi distributions within $\Omega$
are not uniform on $\Xbwd$, but show localization, which is stronger for
larger $\gamma$. 

The same three observations hold for the constructed measures
$\mugh$, where they directly follow from properties $\hone$ and $\htwo$.
The first two observations are implied by conditional invariance.
Note that the integrated weight on $\Omega$ increases with $\gamma$
as $\mugh(\Omega) = 1 - \ue^{-\gamma}$, which follows from Eq.~\eqref{EQ:ci}.
It also implies for the $k$-th preimage of the opening
$\mugh(M^{-k}(\Omega)) = \ue^{-k\gamma}\mugh(\Omega)$,
which agrees with the quantum mechanical analysis \cite{KeaNovPraSie2006}.
For the third observation we explicitly need the saddle distance
in our classical construction,
which follows from property $\htwo$.
Those parts of $\Omega$ with maximal saddle distance $\tcsmax$,
see Fig.~\ref{FIG:TIMES}(d), show the largest enhancement
due to the exponential factor in Eq.~\eqref{EQ:mugh}.
Consequently, regions with smaller saddle distance are less enhanced.
In conclusion we have found a classical explanation for the
localization of resonance eigenfunctions.
In particular, this shows that it is not an interference effect.

Note that our previously proposed measures \cite{KoeBaeKet2015},
which do not depend on $h$, only
resemble the first two observations,
but not the localization effect within the opening.
Thus those measures fail to describe resonance eigenfunctions on
a detailed level.

\begin{figure}[t!]
    \includegraphics{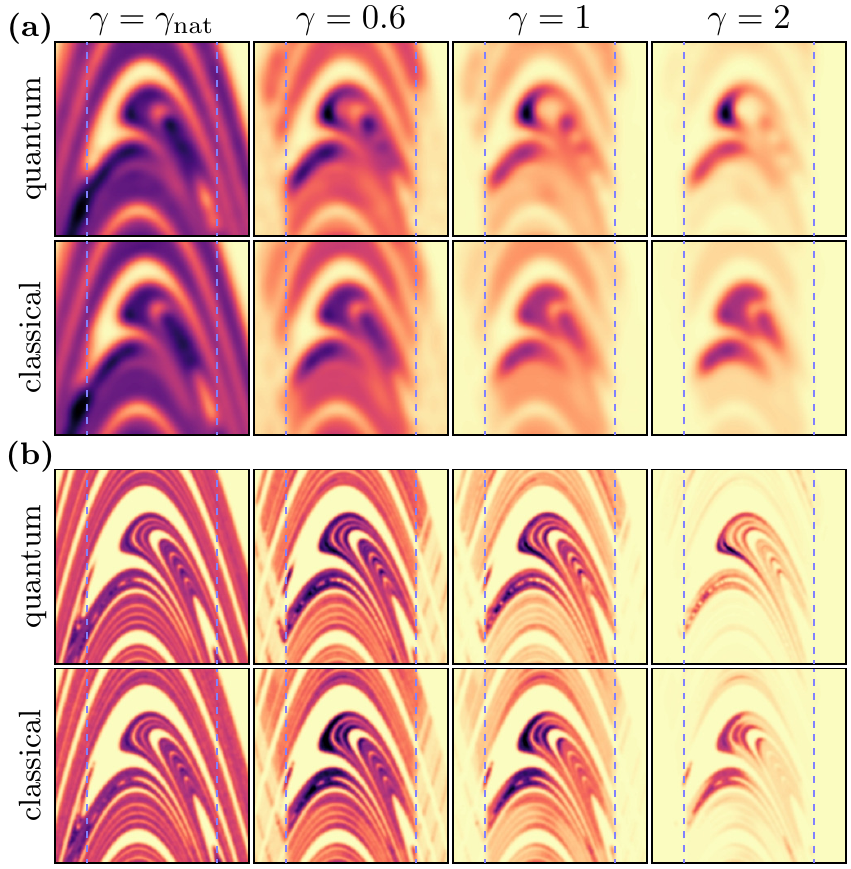}\\
    \caption{Average Husimi distribution of resonance eigenfunctions (top)
        compared to constructed classical measures $\mugh$ (bottom)
        with $\gamma \in \{ \gnat, 0.6, 1, 2 \}$
        for (a) $h=1/1000$ and (b) $h=1/16000$
        on phase-space region $[0.15, 0.45]\times[0.15, 0.45]$.
        Colormap as in Fig.~\ref{FIG:title}, with fixed maximum for
        each $\gamma$ in (a) and in (b).}
    \label{FIG:Zoom}
\end{figure}%
\emph{Dependence on $\gamma$.}---
In Fig.~\ref{FIG:Zoom}(a) we illustrate
quantum (top) and classical (bottom) phase-space distributions zoomed into the
phase-space region $(q,p) \in [0.15, 0.45]\times[0.15,0.45]$ for increasing 
decay rates $\gamma$ starting with $\gnat$ for $h = 1/1000$.
This region is chosen to contain the significant peaks in $\Omega$.
As expected, at the natural decay rate $\gnat$ the Husimi distribution
is almost perfectly resembled by the (smoothed) natural measure $\munat$.
Eigenfunctions with larger $\gamma$ show an increasingly prominent localization.
Classically, this is reproduced using the measures \eqref{EQ:mugh}.
Note that at $\gamma = 2$ also differences between classical and quantum
densities can be seen.
The main peak is sharper and stronger localized quantum mechanically
than for the classical construction.
We attribute this to the chosen simplification using
an integer saddle distance.

\emph{Dependence on $h$.}---%
Figure~\ref{FIG:Zoom}(b) shows the corresponding sequence of plots
for much smaller effective Planck's constant $h=1/16000$.
The eigenfunctions resolve finer structures of
the backward trapped set. %
Again, similarly good agreement between
quantum and classical densities is found.
In particular one observes stronger density variations on $\Xbwd$
in form of arcs, e.g.\ for $\gamma = 1$.
Classically their origin is the increased maximum saddle distance $\tcsmax = 4$
and the finer partition of $\Xbwd$ seen in Fig.~\ref{FIG:TIMES}(e)
especially in the opening.
Furthermore, the sets of maximal saddle distance $\tcsmax$
are similar, see Figs.~\ref{FIG:TIMES}(d) and (e), such that
the localization occurs in a similar region in
Figs.~\ref{FIG:Zoom}(a) and (b).
Again, at $\gamma = 2$ sharper and stronger peaks occur in the
quantum distribution than classically.

While numerically it is not possible to go to much smaller values of the
effective Planck's constant $h$, %
we briefly speculate about the semiclassical limit.
Decreasing $h$ gives a smaller surrounding of
$\saddle$, such that the saddle distance $\sdist$ increases
for all $x\in\Xbwd$, including the maximum $\tcsmax$.
If for decreasing $h$ the difference $\tcsmax - \sdist$ converges,
one can show that
the measures $\mugh$ converge towards a family of $\gamma$-dependent measures
$\mugamma$.
In this case according to the hypothesis a semiclassical convergence of
the eigenfunctions is expected.

If such limit measures $\mugamma$ exist, it is a challenging
question, whether and how they can be calculated directly.
Moreover one would have to test, whether the structure of resonance 
eigenfunctions for finite $h$ is well enough explained by $\mu_\gamma$.
%

\emph{Discussion.}---%
We have shown that the proposed resonance eigenfunction hypothesis
for chaotic systems
reproduces the average phase-space distribution
of resonance eigenfunctions down to scales of order $h$.
In particular the resulting measures $\mugh$
give a classical explanation of the quantum mechanically observed localization.
Small deviations might be improved by more elaborate %
definitions of $\hsaddle$ and the saddle distance $\sdist$,
e.g.\ by considering in the definition of $\hsaddle$ the distance
along the unstable manifold
or by considering continuous saddle distances from a smooth quantum resolved
saddle.
An application of the hypothesis to time-continuous systems,
like open billiards and potential systems, is straightforward.
A future challenge is the application to optical microcavities,
which requires a generalization to partial transmission and reflection.

\begin{acknowledgments}
We are grateful to E.~G.~Altmann, L.~Bunimovich, T.~Harayama, E.~J.~Heller,
S.~Nonnenmacher, and H.~Schomerus
for helpful comments and stimulating discussions, and
acknowledge financial support through the Deutsche Forschungsgemeinschaft
under Grant No. KE 537/5-1.
\end{acknowledgments}

\end{document}